\documentclass[fleqn,usenatbib]{mnras}

\usepackage[T1]{fontenc}
\usepackage{ae,aecompl}

\usepackage{graphicx}	
\usepackage{amsmath}	
\usepackage{amssymb}	

\usepackage{color}
\newcommand{\oiii}[1]{{\ensuremath{\mathrm{O}}}\,\textsc{iii}}
\newcommand{\nii}[1]{{\ensuremath{\mathrm{N}}}\,\textsc{ii}}

\newcommand{\chandra}{{\it{Chandra}} \,}
\newcommand{\ciao}{{\texttt{CIAO}} \,}

\newcommand{\marxx}{{\texttt{MARX}}}

\title[Extended X-ray emission in IC 2497]{Extended X-ray emission in the IC\,2497 - Hanny's Voorwerp system: energy injection in the gas around a fading AGN}

\author[Lia F. Sartori et al.]{Lia F. Sartori$^{1}$\thanks{E-mail:
lia.sartori@phys.ethz.ch}, Kevin Schawinski$^{1}$, Michael Koss$^{1}$, Ezequiel Treister$^{2}$, 
\newauthor W. Peter Maksym$^{3}$, William C. Keel$^{4}$, C. Megan Urry$^{5}$, Chris J. Lintott$^{6}$,
\newauthor O. Ivy Wong$^{7}$\\
$^{1}$Institute for Astronomy, Department of Physics, ETH Z\"{u}rich, Wolfgang-Pauli-Strasse 27, CH-8093 Z\"{u}rich, Switzerland,\\
$^{2}$Departamento de Astronom\'{i}a, Universidad de Concepci\'{o}n, Casilla 160-C, Concepci\'{o}n, Chile,\\
$^{3}$Harvard-Smithsonian Center for Astrophysics, 60 Garden Street, Cambridge, MA 02138, USA,\\
$^{4}$Department of Physics and Astronomy, University of Alabama, Tuscaloosa, AL 35487, USA,\\
$^{5}$Yale Center for Astronomy \& Astrophysics, Physics Department, P.O. Box 208120, New Haven, CT 06520, USA\\
$^{6}$Oxford Astrophysics, Denys Wilkinson Building, Keble Road, Oxford OX1 3RH, UK\\
$^{7}$ICRAR, The University of Western Australia M468, 35 Stirling Highway, Crawley, WA 6009, Australia}

\date{Accepted XXX. Received YYY; in original form ZZZ}

\pubyear{2016}

\begin{document}
\label{firstpage}
\pagerange{\pageref{firstpage}--\pageref{lastpage}}
\maketitle

\begin{abstract}
We present deep \chandra X-ray observations of the core of IC 2497, the galaxy associated with Hanny's Voorwerp and hosting a fading AGN. We find extended soft X-ray emission from hot gas around the low intrinsic luminosity (unobscured) AGN ($L_{\rm bol} \sim 10^{42}-10^{44}$\,erg\,s$^{-1}$). The temperature structure in the hot gas suggests the presence of a bubble or cavity around the fading AGN ($\mbox{{\texttt{E}}$_{\rm bub}$} \sim 10^{54} - 10^{55}$ erg). A possible scenario is that this bubble is inflated by the fading AGN, which after changing accretion state is now in a kinetic mode. Other possibilities are that the bubble has been inflated by the past luminous quasar ($L_{\rm bol} \sim 10^{46}$\,erg\,s$^{-1}$), or that the temperature gradient is an indication of a shock front from a superwind driven by the AGN. We discuss the possible scenarios and the implications for the AGN-host galaxy interaction, as well as an analogy between AGN and X-ray binaries lifecycles. We conclude that the AGN could inject mechanical energy into the host galaxy at the end of its lifecycle, and thus provide a source for mechanical feedback, in a similar way as observed for X-ray binaries.
\end{abstract}




\section{Introduction}\label{sec:intro}

IC\,2497 is a massive ($M \sim 10^{10} M_{\rm \odot}$), nearby ($z = 0.05$) spiral galaxy associated with Hanny's Voorwerp (HV hereafter), an extended emission line region (11 x 16 kpc in projected extent) located at $\sim 20$ kpc in projection from the core of the galaxy and closely matching its redshift. HV was discovered by Hanny van Arkel\footnote{Hanny's Voorwerp means Hanny's object in dutch} a citizen scientist participating in the Galaxy Zoo project (\citealt{Lintott2008,Lintott2009}). The optical spectrum of HV is dominated by [OIII]$\lambda \lambda 4959, 5007 \AA$ emission, and the presence of lines of high ionisation species such as [NeV]$\lambda \lambda 3346, 3426$ and HeII$\lambda 4616$ suggests that the cloud is photoionised by the hard continuum of an active galactic nucleus (AGN) in IC\,2497 rather than star formation (\citealt{Lintott2009}). Moreover, the relatively quiet kinematics (line widths $< 100$ km/s) and low electron temperature ($T_{\rm e} = 13500 \pm 1300 K$) exlude the possibility of ionisation from shocks (\citealt{Lintott2009}). 

Large scale radio observations of the system revealed an extended HI structure ($ M = 8.5 \pm 2.1 \times 10^{10} M_{\rm \odot}$) with irregular kinematics around the southern part of IC\,2497, which suggests a tidal origin (\citealt{Jozsa2009}). These observations also show a kpc-scale structure which may be a jet coming from the AGN. HV lies where the jet meets the HI reservoir and corresponds to a local decrement of HI column density which may be due to photoionisation. At smaller scales, \cite{Rampadarath2010} found a radio AGN in the core of IC\,2497, and a second nuclear source which may be a jet hotspot in the large scale jet reported by \cite{Jozsa2009}.

In order to produce sufficient ionising photons to power the observed [OIII] emission in HV, the quasar in IC\,2497 has to have a bolometric luminosity of at least $L_{\rm bol} = 10^{46}$erg s$^{-1}$. However, the optical nuclear spectrum shows only very weak emission lines, so that the ionising source is classified as LINER or low-luminosity Seyfert galaxy (\citealt{Lintott2009}; \citealt{Keel2012b}). The two possible explanations to reconcile observed and expected emission are 1) the quasar is obscured only along our line of sight but not in the direction of HV, or 2) IC\,2497 hosts a faded AGN. This second option would mean that the quasar dropped in luminosity within the last $\sim 200$ kyr, the travel time needed from the photons to reach the cloud (\citealt{Keel2012b}), but HV still remains lit up because of this travel time. Support for the fading scenario was given by \cite{Schawinski2010b} who analysed the IR data from IRAS and X-ray data from \emph{XMM} and \emph{Suzaku} obtained for IC\,2497. First, as already reported in \cite{Lintott2009} the IR data show no evidence of the reprocessed luminosity expected from an obscured strong quasar. In addition, the \emph{XMM} spectrum is best fitted with an unobscured power-law with photon index $\Gamma = 2.5 \pm 0.7$, as expected from an unobscured AGN, and a component for the hot gas in the galaxy. The quality of the fit is not improved if additional absorption and obscuration are taken into account. This model implies a bolometric luminosity of the AGN $L_{\rm bol} = 4.2 \times 10^{42}$erg s$^{-1}$, which is $\sim 4$ orders of magnitude lower than what expected from HV. In addition, \emph{Suzaku} observations show only a marginal detection with the 15-30 keV PIN camera (hard X-ray detector). If these counts are real, this would mean that tha AGN is strongly obscured and we are observing the reemission instead of the AGN power-law, but the obtained luminosity would still be $\sim 2$ orders of magnitude below what expected from HV. Also the \emph{HST} observations presented by \cite{Keel2012b} argue against the hypothesis of a luminous, strongly obscured AGN. WFC3 images show a complex dust structure near the nucleus of the galaxy, but the view of the nucleus is not hindered by absorbing features, so that obscuration cannot be the explanation of the lack of strong AGN features. Moreover, there is no high-ionisaion gas near the nucleus, suggesting that the current radiative output from the AGN is low.

The observations described above suggest that the quasar in IC\,2497 dropped in luminosity by at least 2 orders of magnitude in the last $\sim 10^5$ years. The system composed by HV and its galaxy is therefore a great laboratory to study AGN variability on previously inaccesible timescales. The analysis of the galaxy allows us to study the present state of the AGN and the AGN-host galaxy interaction in this ``post quasar" phase, while HV provides information about the past state of the AGN. Finally, the distance between the two provides information about timescales.


In this paper we present the analysis of new, deep \chandra X-ray observations of IC\,2497. The paper is organised as follows. In Section \ref{sec:data} we describe the X-ray observations and the data reduction. \textcolor{black}{In Section \ref{sec:im_spec} we describe the performed imaging and spectroscopic analysis. Because of the good spatial resolution reached by Chandra, this analysis allows us not only to measure the overall X-ray flux from the galaxy, but also to (partially) resolve its emission.} In Section \ref{sec:disc} we discuss the results and possible consequences for the AGN-host galaxy interaction as well as the analogy bewteen AGN and X-ray binaries.

\section{Observations}\label{sec:data}

We observed IC\,2497 with the \chandra Advanced CCD Imaging Spectrometer (ACIS) on 2012 January 8 (ObsID 13966, 59.35 ks) and 2012 January 11 (ObsID 14381, 53.4 ks; PI Kevin Schawinski, Cycle 13). We took the data with the S-array (ACIS-S) in very faint (VFAINT) time-exposure (TE) mode. 
We performed a standard data reduction starting from the level 1 event files using the \ciao 4.7 software \citep{Fruscione2006} provided by the \chandra X-ray Center (CXC). We ran the script {\texttt{chandra$\_$repro}} with the latest CALDB 4.6.8 set of calibration files, applied subpixel position using the Energy Dependent Subpixel Event Repositioning algorithm ({\texttt{pix$\_$adj$=$EDSER}}; \citealt{Tsunemi2001}; \citealt{Li2003, Li2004}), and flagged the background events most likely associated with cosmic rays by running the task {\texttt{acis$\_$process$\_$events}} ({\texttt{check$\_$vf$\_$pha$=$yes}} in {\texttt{chandra$\_$repro}}). We analysed the background light curves with the {\texttt{CIAO}} routine {\texttt{lc$\_$sigma$\_$clip()}} and found no interval of unusually strong background flaring. Because of the low count rate the observations show no evidence of pileup.

\textcolor{black}{In order to correct for the offset in the relative astrometry of the two observations, we first rebinned the reprocessed event files to 1/4 pixel resolution (0.125 arcsec) using the {\texttt{CIAO dmcopy}} routine with binning factor 0.25. We then ran the {\texttt{CIAO dmstat}} routine on the rebinned images to get the centroid of the emission from IC\,2497.} In the following we will consider the centroid as the centre of the emission in the respective observation.

We extracted the spectra and generated the response files using the \ciao tool {\texttt{specextract}} (see Section \ref{sec:im_spec} for more details about the extraction regions). For the background estimation we considered three source-free circular regions with a 5 arcsec radius $\sim$ 20 arcsec away from the source. We then grouped each spectrum with a minimum of 3 counts per bin using the {\texttt{Heasoft}} tool {\texttt{grppha}}.

We simulated the PSF images needed for the analysis using {\texttt{MARX 5.1}} \citep{Davis2012}. First, we created a model of the emission using {\texttt{Xspec}} and {\texttt{Sherpa}} (see Section \ref{sec:im_spec} for details about the assumed models). We then used {\texttt{MARX}} to run the raytrace simulation and project the ray-tracings onto the ACIS-S detector. Since we are interested on the sub-pixel regime we ran the simulations with {\texttt{pixadj=EDSER}} and {\texttt{AspectBlur=0.19}} arcsec (telescope pointing uncertainty). We also included the telescope dithering ({\texttt{DitherModel=INTERNAL}}, {\texttt{DitherAmp$\_$Ra=8}}, {\texttt{DitherAmp$\_$Dec=8}}) and corrected for SIM drift and offset\footnote{For details on simulating a PSF image using {\texttt{MARX}} see \url{http://cxc.harvard.edu/chart/threads/marx_sim/}}.

\section{Analysis}\label{sec:im_spec}

\subsection{Radial profile and source extension}



\begin{figure*}
\includegraphics[scale=0.64]{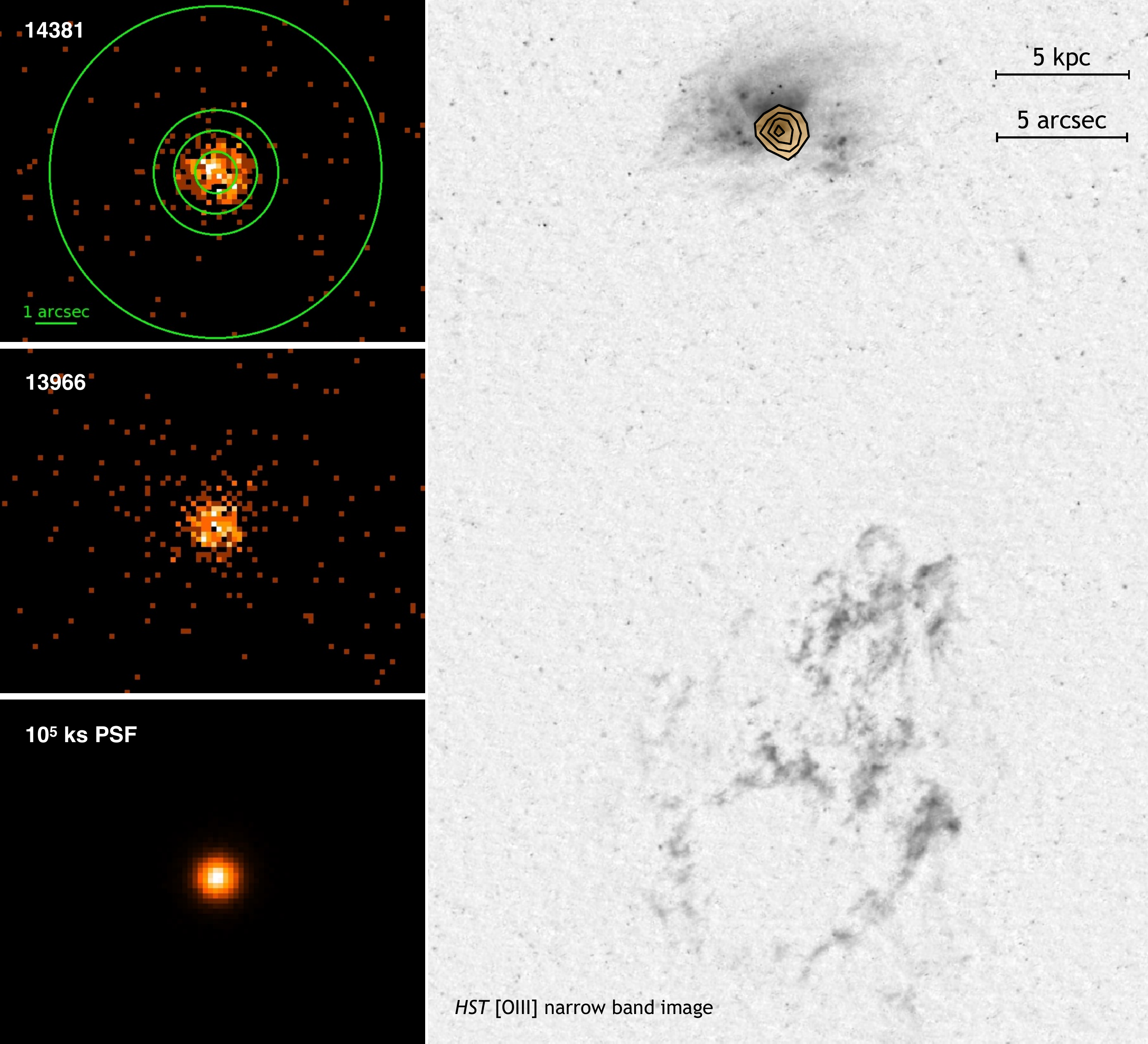}
\caption{Left: \chandra X-ray images for ObsID 14381 (53.4 ks), ObsID 13966 (59.35 ks), and simulated 10$^5$ ks PSF. The green annuli show the regions from which we extract the spectra and compute the brightness profile. The spatial scale is the same for each image. The colour scale is linear (0-5 counts for the observations, 0-1800 counts for the 10$^{5}$ ks PSF). Right: \chandra X-ray contours overplotted on the {\it{Hubble}} [OIII] narrow band image of IC\,2497 (top) and Hanny's Voorwerp (bottom; from \protect\citealt{Keel2012b}).}
\label{fig:images}
\end{figure*}

\begin{figure}
	\includegraphics[width=\columnwidth]{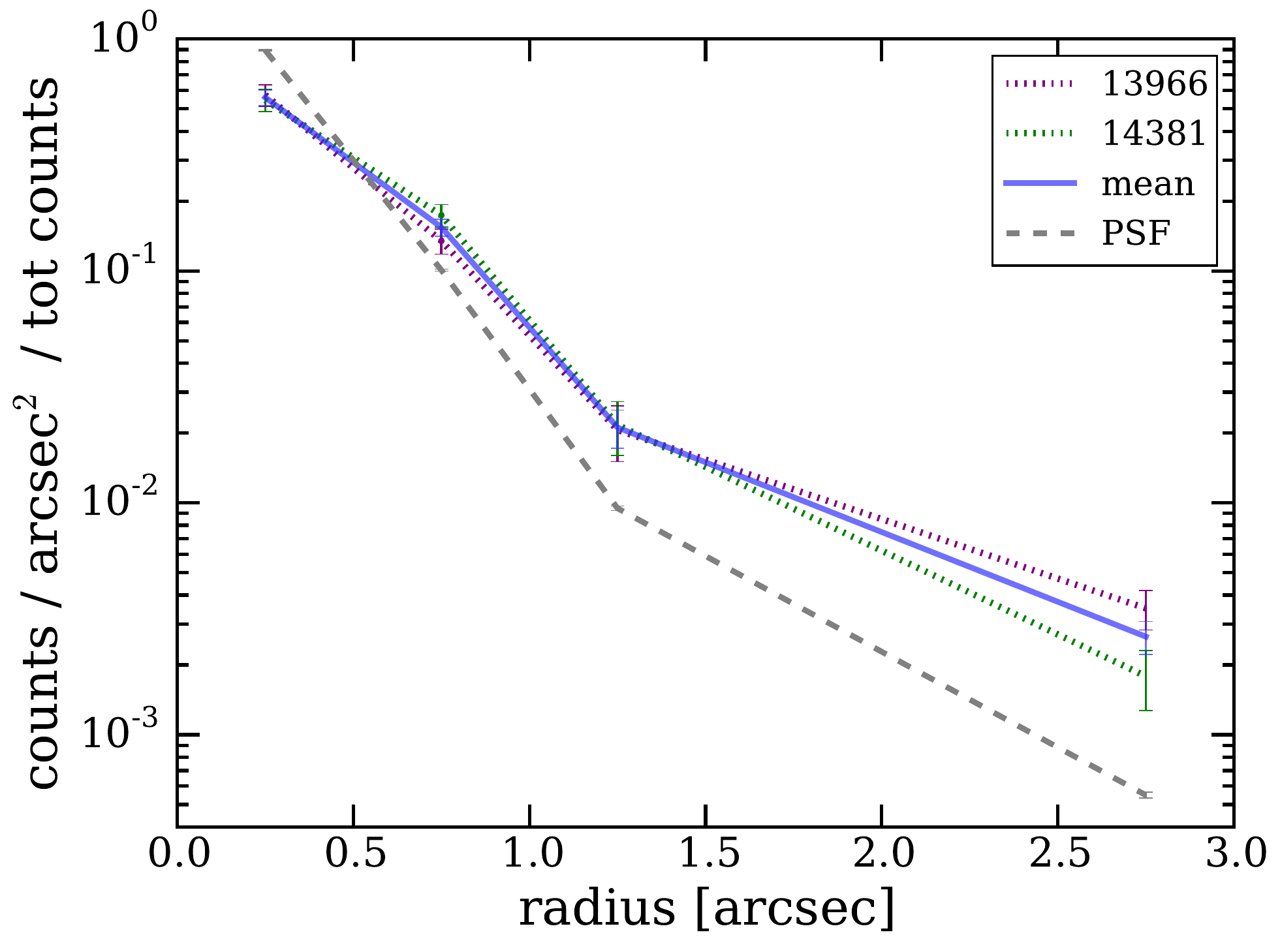}
    \caption{Normalized surface brightness profiles of the \chandra images and of the simulated PSF. ObsID 13966 and 14381 are shown in purple and green, respectively, while the blue line shows the mean between the two observations. The profiles of the two observations are consistent within 1 $\sigma$ (2 $\sigma$ in the last annulus). The gray line shows the profile of the simulated PSF. The PSF profile is steeper toward the center than the source's profile. This confirms that the source is extended.}
    \label{fig:br_prof}
\end{figure}

We compute the 0.5 - 6.0 keV background-subtracted surface brightness profiles of the two observations starting from the reprocessed event files. We use the \ciao tool {\texttt{dmextract}} to extract the counts in four concentric annuli centered on the centroid of each exposure, and with outer radius 0.5, 1.0, 1.5 and 4.0 arcsec (Fig. \ref{fig:images}, first panel). For the background we consider a source-free annulus with internal radius 4 arcsec and external radius 7 arcsec. The obtained radial profiles, normalized to the total number of counts in a 4 arcsec aperture, are shown in Fig. \ref{fig:br_prof}. The error bars correspond to the 1 $\sigma$ confidence interval (68.23$\%$). For the annuli with more than 25 counts (N > 25) we compute the 1 $\sigma$ error of the total counts (not background subtracted) using Gaussian statistic, $\sigma = \sqrt{N}$. For N $\leq$ 25 we use the Gehrels approximation $\sigma = 1 + \sqrt{N + 0.75}$ (\citealt{Gehrels1986}; this actually corresponds to the 84.13$\%$ upper limit for a Poisson distribution). The profiles of the two observations are consistent within 1 $\sigma$ (2 $\sigma$ in the last annulus), showing no evidence of variability.

\begin{figure}
	\includegraphics[width=\columnwidth]{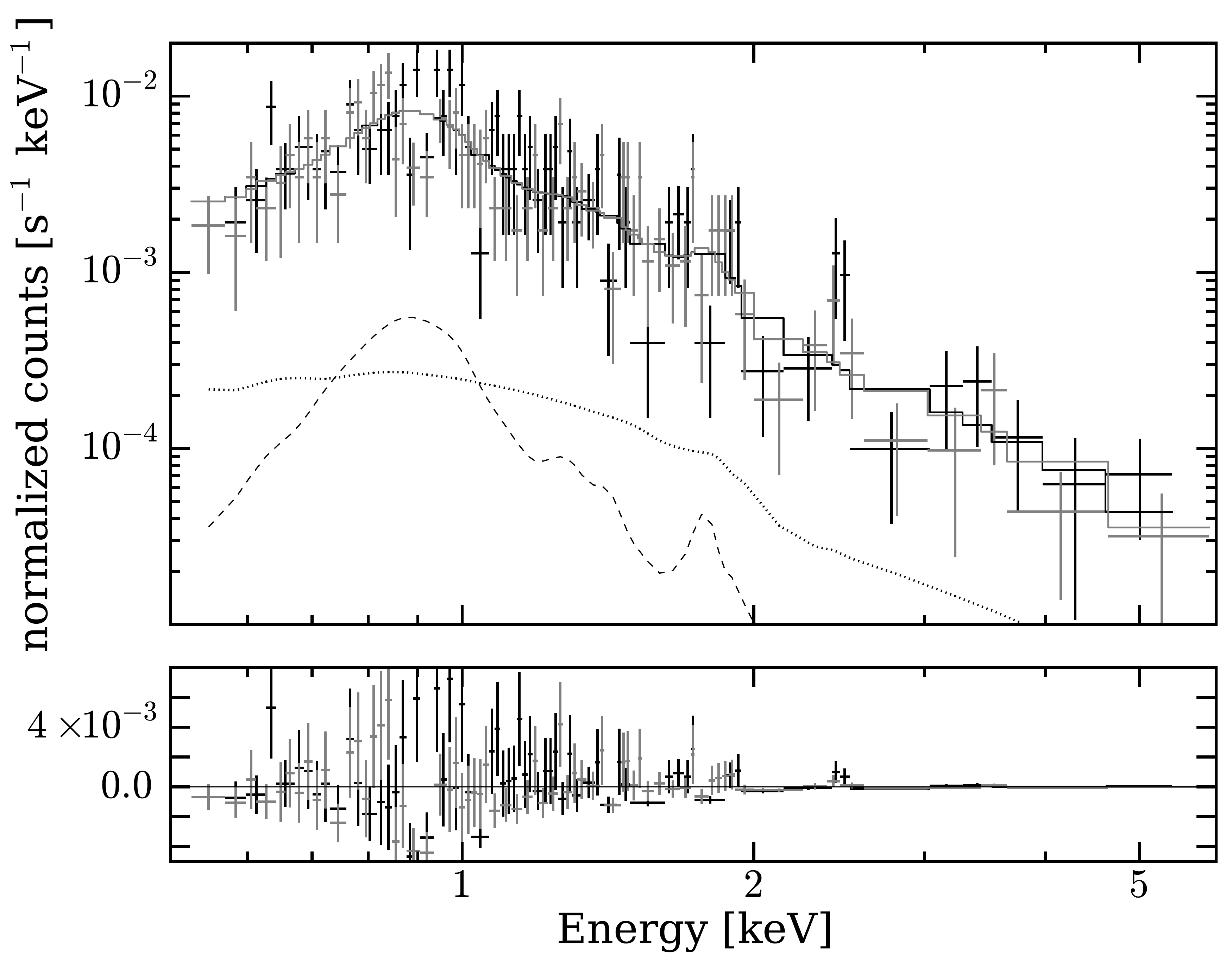}
    \caption{\chandra 0.5 - 6.0 keV X-ray spectrum of IC\,2497 (total emission, r = 4 arcsec) for ObsID 14381 (black) and ObsID 13966 (grey). The spectra are binned so that each bin contains at least 3 counts. The lines correspond to the best fit model {\texttt{phabs}}$\times$({\texttt{zpow}}+{\texttt{APEC}}), where {\texttt{zpow}} corresponds to the AGN emission and {\texttt{APEC}} to the emission from a hot diffuse gas (see text). The absorption is fixed to the Galactic value. The residuals of the fit are shown in the bottom panel. \textcolor{black}{The dotted and dashed lines correspond to the \texttt{phabs}$\times${\texttt{zpow}} and \texttt{phabs}$\times${\texttt{APEC} models, respectively, both normalised by a factor 10.}}}
    \label{fig:spec_tot}
\end{figure}

The emission from an AGN is expected to be point-like. However, the radial profile suggests that the emission in IC\,2497 is extended. Moreover, the spectrum extracted from a 4 arcsec aperture is best-fit with a combination of two models: a power-law model for the AGN emission, and a collisionally ionised diffuse gas model (APEC) representing the hot gas in the galaxy, which can be expected to be extended (Fig. \ref{fig:spec_tot}, Table \ref{tab:par}). This is true both performing a simultaneous fit of the two observations, or considering them separately (the best-fit parameters are always consistent within the 90$\%$ confidence level, see more details about spectral fit in Section \ref{sec:spec}). In order to understand if the observed extension is physical, and not only due to the PSF, we compare the radial profile of the observations with the expected PSF profile. We simulate the PSF image as described in Section \ref{sec:data}, starting from the fit to the total emission. We first create a model of the total emission using {\texttt{Xspec}} and {\texttt{Sherpa}}, and the spectral model described above, where we also consider Galactic absorption ({\texttt{phabs}}$\times$({\texttt{zpow}}+{\texttt{APEC}})), and then use {\texttt{MARX}} to run the ray trace simulation and project the ray-tracings onto the ACIS-S detector. In order to get a good sampling of the PSF, and thus understand its analytic shape, we increase the exposure time to 10$^{5}$ ks. We compute the PSF brightness profile in the same way as described for the real observations. Since the PSF profile should in principle depend on the input spectrum and on the source position on the ACIS-S detector we repeat the simulations for both observations separately. However, due to the similarity of model spectra and positions (both observations are nearly on-axis) the normalized PSF brightness profiles are the same. 

Fig. \ref{fig:br_prof} shows the comparison between the normalized brightness profile of the source (two observations and mean value) and of the PSF. The PSF profile is steeper toward the center than the source's profiles. This confirms that the source is extended.

\subsection{Spectral fit and temperaure profile}\label{sec:spec}

For both observations we use {\texttt{specextract}} to extract the spectrum of the total X-ray emission in IC\,2497 from an aperture with r = 4 arcsec centered on the centroid. \textcolor{black}{We fit the spectra in the 0.5 -- 6.0 keV energy range with {\texttt{Xspec}} using Cash statistic \citep{Gehrels1986} as fit statistic, and Chi-Square statistic as test statistic}.  Since there is no variability in the observations and between the observations at the 90$\%$ level we fit the spectra from the two data sets simultaneously. The source is best fitted as a sum of two components: a power-law model representing the AGN emission ({\texttt{zpow}} model, $\Gamma = 2.33^{+0.15}_{\rm -0.16}$), and a model for the hot gas emission at the redshift of the host galaxy, with abundance fixed at solar value ({\texttt{APEC}} model, $kT = 0.953^{+0.048}_{\rm -0.054}$ keV)\footnote{\textcolor{black}{Since IC 2497 is a massive galaxy, the assumption of solar abundance should be appropriate. However, we also performed the fits considering metallicities of 0.2 and 0.5 solar,  and the differences in the measured temperatures are only at the $<10\%$ level. This is true both for the total emission and for the annuli described in the next paragraphs}}. The absorption component ({\texttt{phabs}}) is fixed at the Galactic value ($N\rm _H = 1.31 \times 10^{20}$ cm$^{-2}$, as given by the Colden Calculator with the NRAO dataset\footnote{\url{http://cxc.harvard.edu/toolkit/colden.jsp}}). All the best fit parameters are listed in Table \ref{tab:par}. The power-law component is consistent with a low luminosity AGN with $L_{\rm 2-10 keV} = 5.8 \times 10^{40}$\,erg\,s$^{-1}$.\\

In order to understand the properties of the extended emission we attempt a spatially resolved spectral analysis. For each observation we use {\texttt{specextract}} to extract the spectrum from a circular aperture of radius 0.5 arcsec centered on the centroid, and 3 annuli with outer radius 1.0, 1.5 and 4.0 arcsec. Since we will account for the PSF during the fit, as described in details in the following, we set {\texttt{correctpsf=no}}. We then fit the spectra in the same way as described for the total emission. The best fit parameters are listed in Table \ref{tab:par}, while the fitted spectra are shown in appendix \ref{app:fit}.


The AGN emission is expected to be point-like and the {\texttt{zpow}} component should in principle be present in the nuclear spectrum only. However, because of the PSF some AGN contribution is present also in the external annuli. We therefore perform the following analysis, where we fix the AGN component in the fit such that the normalization and the power-law slope are PSF-dependent, and let only the APEC model free to vary. First, we simulate a PSF in the same way as described in Sections \ref{sec:data} and \ref{sec:im_spec} but for a {\texttt{zpow}} model only, extract the spectra from the annuli considered for the analysis, and fit them using the response files (RMF and ARF files) obtained for the real spectra\footnote{in this way we treat the simulated spectra as similar as possible to the real spectra. We repeated the analysis also using simulated RMF and ARF files, but the fit parameters do not differ significantly.}. Since the number of counts in the simulated spectra is high, we group them with a minimum of 25 counts per bin and use Gaussian statistic. In each annulus the spectrum is consistent with a power-law at the redshift of the galaxy ({\texttt{zpow}}). However, since the PSF also depends on the energy, the power-law slope ({\texttt{PhoIndex}}) varies with radius, as described in Table \ref{tab:par}. We therefore fit the real spectra with the {\texttt{phabs}}$\times$({\texttt{zpow}}+{\texttt{APEC}}) model described above, where for each annulus we fix the power-law slope at the value obtained from the PSF simulation. In order to fix the power-law normalisation, we first fit the total emission (0.0 -- 4.0 arcsec), and assume that in each annulus the normalisation scales in the same way as seen in the PSF simulation:

\begin{equation}
\mbox{{\texttt{norm$\_$pl}}$_{\rm i}$} = \frac{\mbox{norm$\_$PSF$_{\rm i}$}}{\mbox{\mbox{norm$\_$PSF$_{\rm tot}$}}} \times \mbox{{\texttt{norm$\_$pl}}$_{\rm tot}$}
\end{equation}

where $\mbox{{\texttt{norm$\_$pl}}$_{\rm i}$}$ and $\mbox{{\texttt{norm$\_$pl}}$_{\rm tot}$}$ are the normalisation of the power-law component in the real spectra extracted from the $i$-te annulus and from the total aperture (0.0-4.0 arcsec), while $\mbox{norm$\_$PSF$_{\rm i}$}$ and $\mbox{norm$\_$PSF$_{\rm tot}$}$ are the same but for the PSF simulation.

\begin{figure}
	\includegraphics[width=\columnwidth]{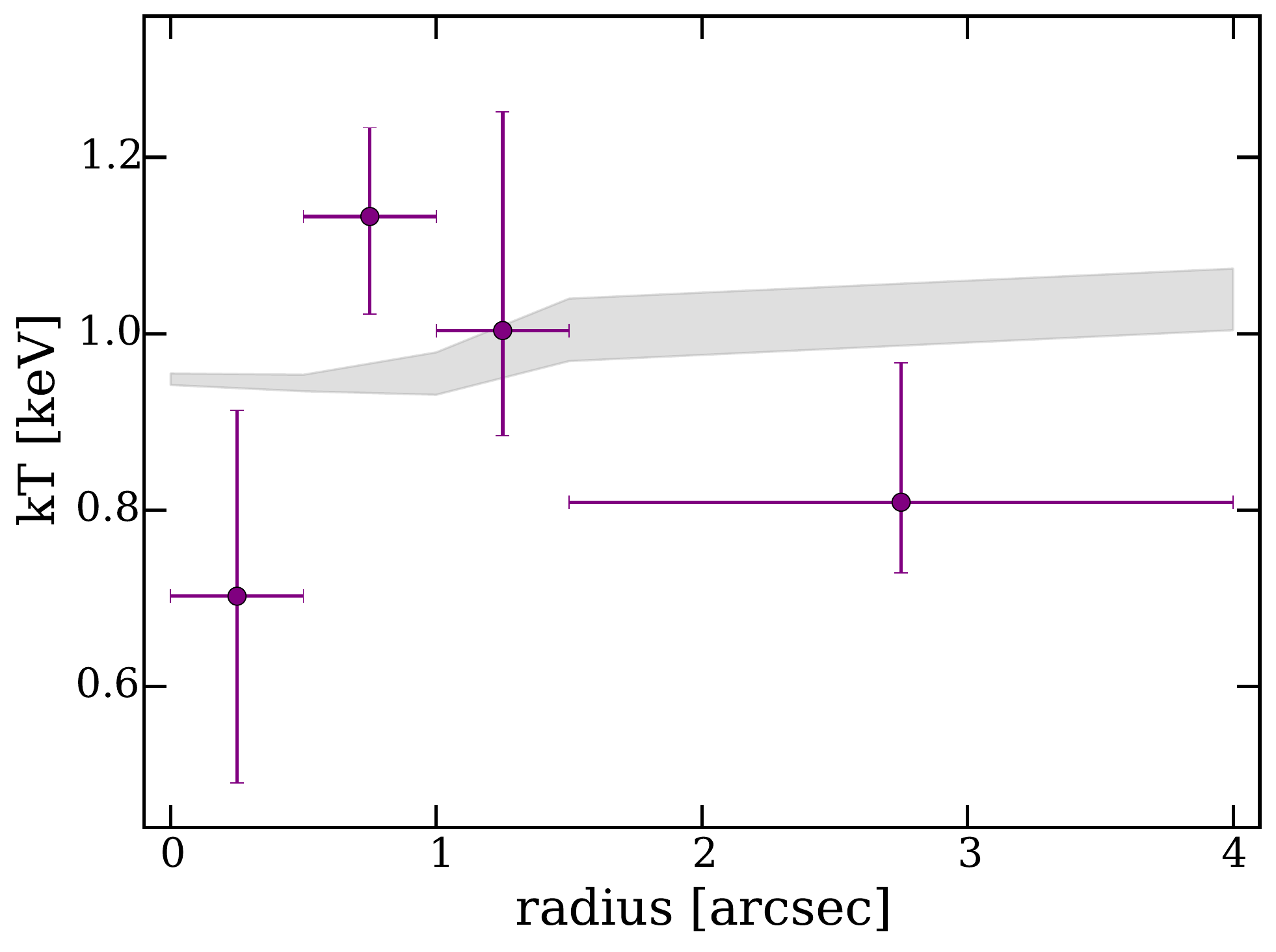}
    \caption{Projected temperature profile from the spatially resolved spectral analysis. Each bin corresponds to one annulus defined in Fig. \ref{fig:images}. The purple points correspond to the real data. There is a hint of a temperature gradient from the center to the first two annuli with a statistical significance above 1 $\sigma$, which could indicate the presence of a bubble powered by the central AGN. The gray shaded area corresponds to the temperature profile measured in the simulated PSF (see Section \ref{sec:spec} for more details). This temperature does not show the gradient seen in the real data, which confirms that the gradient is not due to the PSF.}
    \label{fig:t_gradient}
\end{figure}

In Fig. \ref{fig:t_gradient} we show the projected temperature profile (from the {\texttt{APEC}} parameter {\texttt{kT}}). We find a hint of increasing temperature from the center toward the first and second annuli with a statistical significance above 1 $\sigma$. We discuss possible implications of the found temperature gradient in Section \ref{sec:disc}. In order to understand if this gradient is due to systematics in the fits we perform a Monte Carlo simulation: first we use the {\texttt{Xspec}} function {\texttt{fakeit}} to simulate spectra similar to the observed ones (similar power-law slope and normalisation) with a given temperature profile. We then fit the simulated spectra using the same procedure as for the real data. Our simulations show that we can recover the input profile within 1 $\sigma$, and thus the gradient is not due to systematics. We also investigate the possibility that the gradient is artificially introduced by our fitting procedure. We repeat the fit of the real data without fixing the power-law slope and normalisation. Also in this case we find a temperature gradient consistent at the $90\%$ confidence level with the one showed in Fig. \ref{fig:t_gradient}. Finally, we investigate the possibility that the temperature gradient is due to a PSF effect. We simulate a PSF with the same model as the total emission ({\texttt{phabs}}$\times$({\texttt{zpow}}+{\texttt{APEC}})) and measure the temperature in the same annuli as used for the real source. As we can see in Fig. \ref{fig:t_gradient} (grey shadow), the measured temperature does not show the gradient seen in the real data, what again confirms that the gradient is not due to the PSF.

The temperature increase showed in Fig. \ref{fig:t_gradient} could be symmetric, or due to a feature in a preferred direction. In order to investigate these possibilities we repeat the analysis dividing the first annulus (0.5 -- 1.0 arcsec) into 4 semi annuli of 90 deg each. The results are listed in Table \ref{tab:par}. Because of the lower number of counts in each region, these fits are less reliable than the one to the total annulus. However, the temperature gradient seems to be present in each direction, although it is more important in the southern part. This seems to rule out the possibility of a single feature with significantly higher temperature.

It is important to note though that the low number of counts in our spectra do not allow us to deproject the emission from the different shells, so that the reported gas temperatures and normalisations are projected values along the line of sight. A deeper $\sim 1$ Ms \chandra exposure will allow us to obtain a minimum of 200 counts in all region of interest, and thus properly deproject the emission and carefully measure the intrinsic gas parameters.



\begin{table*}
 \begin{tabular}{lccccccc}
  \hline
  Region & Net Counts & PL slope & PL norm & Temperature & {\texttt{APEC}} norm\\
   & & & [$10^{-7}$] & kT [keV] & [$10^{-7}$]\\
  \hline
  Total emission (0.0-4.0 arcsec) & 620 & $2.33^{+0.15}_{\rm -0.16}$ & $67.486^{+9.453}_{\rm -9.162}$ & $0.953^{+0.048}_{\rm -0.054}$ & $38.735^{+6.558}_{\rm -6.381}$\\
  \hline
  Nucleus (0.0-0.5 arcsec) & 262 & 2.47 & 46.708 & $0.702^{+0.211}_{\rm -0.212}$ & $5.193^{+1.941}_{\rm -1.823}$\\[0.2cm]
  Annulus 1 (0.5-1.0 arcsec) & 221 & 2.31 & 15.387 & $1.133^{+0.100}_{\rm -0.111}$ & $21.143^{+4.293}_{\rm -3.974}$\\[0.2cm]
  Annulus 2 (1.0-1.5 arcsec) & 52 & 1.89 & 2.019 & $1.003^{+0.248}_{\rm -0.120}$ & $5.023^{+1.077}_{\rm -1.130}$\\[0.2cm]
  Annulus 3 (1.5-4.0 arcsec) & 84 & 1.65 & 1.213 & $0.809^{+0.158}_{\rm -0.080}$ & $10.013^{+1.425}_{\rm -1.315}$\\
  \hline
  Semi-ann. 1 (45-135 deg) & 54 & 2.28 & 3.709 & $1.064^{+0.249}_{\rm -0.123}$ & $5.185^{+2.893}_{\rm -1.277}$\\[0.2cm]
  Semi-ann. 2 (135-225 deg) & 56 & 2.37 & 3.881 & $0.920^{+0.105}_{\rm -0.139}$ & $5.316^{+1.343}_{\rm -1.180}$ &  \\[0.2cm]
  Semi-ann. 3 (225-315 deg) & 55 & 2.27 & 3.709 & $1.575^{+und.}_{\rm -0.343}$ & $7.933^{+7.583}_{\rm -3.163}$\\[0.2cm]
  Semi-ann. 4 (315-405 deg) & 55 & 2.29 & 3.886 & $1.283^{+0.386}_{\rm -0.185}$ & $7.401^{+4.406}_{\rm -2.173}$\\
  \hline
 \end{tabular}
 \caption{Best fit parameters for each region considered in this analysis (nucleus, three annuli, four semi-annuli). The assumed model is {\texttt{phabs}}$\times$({\texttt{zpow}}+{\texttt{APEC}}). In the nucleus all the fit parameters are free. The power-law slope and normalisation are fixed according to the PSF simulation (see text for details). Abundance is fixed at Solar value and absorption at Galactic value ($N\rm _H = 1.31 \times 10^{20}$ cm$^{-2}$).}
 \label{tab:par}
\end{table*}

\section{Discussion}\label{sec:disc}

We detect extended soft X-ray emission in the nucleus of IC\,2497, in addition to the point-like power-law emission expected from an AGN. We associate this emission with a hot, diffuse gas. The projected temperature profile (Fig. \ref{fig:t_gradient}) shows a resolved cold center. The temperature gradient may be interpreted as a bubble or cavity in the hot gas which is kinematically inflated by the AGN. Alternatively, the temperature gradient could indicate a shock front, e.g. from an AGN-driven superwind. The different hypothesis, and the consequences for the AGN-host galaxy interaction, are described in the following subsections.

The current analysis also supports the idea that IC\,2497 hosts a fading AGN, as suggested by \cite{Lintott2009}, \cite{Schawinski2010b} and \cite{Keel2012b}. In fact, by scaling the quasar spectral energy distribution (SED) templates from \cite{Elvis1994} to match the measured X-ray luminosity, $L_{\rm 2-10 keV} = 5.8 \times 10^{40}$\,erg\,s$^{-1}$, we obtain a bolometric luminosity $L_{\rm bol} \sim 10^{42}$\,erg\,s$^{-1}$, which is $\sim 4$ orders of magnitude below what is needed in order to power Hanny's Voorwerp (\citealt{Lintott2009}; \citealt{Schawinski2010b}). This result is consistent with what reported by \cite{Schawinski2010b}. As reported by \cite{Schawinski2010b}, if the AGN is compton thick, this would mean that it is strongly obscured and we are observing the reemission instead of the AGN power-law, and the bolometric luminosity would be $L_{\rm bol} \sim 10^{44}$\,erg\,s$^{-1}$ (see section \ref{sec:intro}). This is still $\sim 2$ orders of magnitude below what is needed in order to power Hanny's Voorwerp. Hard X-ray observation from {\emph{NuSTAR}} are now crucial in order to understand if the AGN is obscured and measure the real magnitude of the shutdown.

\subsection{Hot gas in IC\,2497}
The projected temperature observed in the gas in IC\,2497 is in the range $kT \approx 0.7 - 1.1$ keV ($T \approx 0.8 - 1.2 \times 10^{7}$K). This relatively high themperature may indicate that the emission from the hot gas is not due to photoionisation, as seen in other Seyfert galaxies (e.g. \citealt{Bianchi2006}, \citealt{Greene2014}). On the other hand, the projected temperature measured in the center and in the external annulus are consistent with what found in ULIRGs (\citealt{Grimes2005}). The high projected temperature, especially in the two central annuli, may be an indication that we are observing a shock front, e.g. from an AGN-driven superwinds. Models of AGN-driven superwinds suggest a temperature of $T \approx 10^{10} - 10^{11}$K for the inner reverse shock and $T \approx 10^{7}$K for the forward shock, and the expanding medium can persist also after the AGN has switched off (\citealt{Zubovas2012}). The observed temperatures are consistent with the forward shock, but too low for the inner shock. Is the AGN in IC\,2497 too faint to produce such a high temperature? Or is the observed shock front a residual of a wind driven by the AGN during the past quasar phase? 
Deeper X-ray observations (leading to X-ray spectra with more counts) will allow us to deproject the emission and carefully constrain the intrinsic temperature, density and X-ray luminosity of the gas in each shell around the AGN. This will allow us to estimate the velocity of the shock and the ratio between X-ray and mechanical luminosity (e.g. \citealt{Chu1990}), and therefore probe the superwind hypothesis and constrain between the two scenarios.


Another possible interpretation of the observed projected temperature gradient is that the AGN is, or was, blowing away the hot gas around it by injecting kinetic energy, therefore creating a bubble or cavity. If the hypothesis of a bubble inflated by a fading AGN is confirmed, this would give us new insights into the AGN-host galaxy interaction in the ``post-quasar" phase, as described below. It is important to note though that the low number of counts in our spectra do not allow us to deproject the emission from the different shells, so that the reported gas temperatures and normalisations are projected values along the line of sight. A deeper $\sim 1$ Ms \chandra exposure will allow us to obtain a minimum of 200 counts in all region of interest, and thus properly deproject the emission and carefully measure the gas parameters. In this way we will be able to further test the bubble hypothesis.   

%

\subsection{Energy in the bubble}
In order to understand the effect of the AGN on the surrounding medium we need to estimate the energy in the bubble, i.e. the energy required to create it. The uncertainties in this calculation are quite high due to the uncertainties in the spectral fitting and the unknown shape of the bubble and hot gas. Therefore, the obtained energy should be seen as an order of magnitude estimation. One possible way to approach the problem is to consider an analogy with the cavities observed in clusters (e.g. \citealt{Birzan2004}, \citealt{Rafferty2006}, \citealt{McNamara2006}, \citealt{Vantyghem2014}). In this case the energy needed to create the bubble can be espressed as:
\begin{equation}
\mbox{{\texttt{E}}$_{\rm bub}$} = \frac{\gamma}{\gamma - 1} p \mbox{{\texttt{V}}$_{\rm bub}$}
\end{equation}

where $p$ is the pressure of the gas surrounding the bubble, $\mbox{{\texttt{V}}$_{\rm bub}$}$ is the bubble's volume, and $\gamma$ is the heat capacity ratio of the gas inside the bubble ($\gamma = 4/3$ for a relativistic gas, $\gamma = 5/3$ for a non-relativistic, monoatomic gas). For our estimation we consider a spherical bubble with radius $r = 1$ arcsec, corresponding to $r = 987$ pc at the redshift of the source. The pressure $p$ can be expressed as:

\begin{equation}
p = n (kT)
\end{equation}

where $n$ and $kT$ are the number density and temperature of the gas outside the bubble, as derived from the spectral fit to the third annulus (1.5-4.0 arcsec). For a fully ionised gas with 10$\%$ He we can assume $n_{\rm e} \sim n_{\rm H}$ and $n \sim 2n_{\rm e}$, so that $n$ can be derived from the \texttt{APEC} normalisation:
\begin{equation}
\mbox{{\texttt{norm}}} = \frac{10^{-14}}{4 \pi \texttt{D}_{\rm A}{}^2 (1+z)^2} \int n_{\rm e} n_{\rm H}\,dV 
\end{equation}

where we consider the normalisation obtained in the third annulus, while the volume corresponds to a napkin ring for a sphere with $r$ = 4 arcsec and a cylinder with $r= 1.5$ arcsec (i.e. the volume of the external shell which is projected into the third annulus). In this way we obtain $n \approx 0.02$cm$^{-3}$ and $p \approx 2.5 \times 10^{-11}$erg cm$^{-3}$, so that the energy in the bubble is $\mbox{{\texttt{E}}$_{\rm bub}$} \sim 10^{54} - 10^{55}$ erg. 

Now we will carry out comparisons to test what these numbers can tell us about the bubble itself and the effect of the AGN on the host galaxy. The energy in the bubble is $\sim 3 - 4$ orders of magnitude higher than what expected from a supernova explosion, so that the bubble cannot be inflated by supernovae. \cite{Keel2012b} found a ring of ionised gas on the northern part of the nucleus of IC\,2497. The ring has a radius $ r \approx 250$ pc which can be seen in both \emph{HST} H$\alpha$ and [OIII] continuum images, and with a kinematic age $< 7 \times 10^5$\,yr (see Figure 5 in \citealt{Keel2012b}). The kinetic energy in this ring is $\sim 3 \times 10^{51}$ erg, $\sim 3 - 4$ orders of magnitude lower than what found in our bubble, and also the spatial extent is different. This is an indication that additional components need to be considered, probably happening at the same time, near the nucleus of IC\,2497. Finally, the binding energy of IC\,2497 is $\sim 10^{60}$ erg, which means that the present-day AGN is not able to blow away the whole ISM. However, \cite{Schawinski2015} recently suggested that AGN ``flicker" on and off 100-1000 times alternating high-Eddington bursts and lower-Eddington troughs points, with a typical AGN phase lasting $\sim 10^5$ yrs. If what we observe in IC\,2497 is typical to each burst, i.e. the AGN injects $\sim 10^{55}$ erg in each cycle, the sum of the energy injected in each cycle could start having some effect on the gas reservoire (although alone it would probably be insufficient to totally unbind it). Finally, if we asume that the bubble is inflated by the fading AGN after the quasar drop in luminosity (i.e. in the last $\sim 2 \times 10^{5}$ yr), the AGN should inject at least $\sim 10^{42}$ erg/s in kinetic form, and the bubble should expand with a velocity $v_{\rm exp} \sim 5000$ km/s.

\subsection{Accretion state changes, AGN feedback and analogy with X-ray binaries}
As explained above, one possible interpretation of the bubble is that it is inflated by the fading AGN after the quasar drops in luminosity, and we are witnessing a state change in the accretion mode of the quasar. After dropping in accretion rate, the low-Eddington AGN moved from a radiative mode to a kinetic mode, and now puts out part of its energy in kinetic form. In this way, it can do a feedback on the galaxy, e.g. by inflating a bubble in the surrounding gas or launching the radio jets. The hypothesis of a state change in IC 2497 is consistent with the analysis done by \cite{Keel2012b}. In fact, the \emph{HST} STIS spectra and images indicate the presence of an outflow from the nucleus which suggests that some of the energy output from the AGN switched from radiation to kinetic form.

The idea of a bubble inflated by an AGN undergoing a state change fits naturally into the general picture of unifying BH accretion physics from X-ray binaries (XRB) to AGN. Rapid state transitions, on timescales of days to weeks, are often observed in XRB, and are connected to a change in radiative output (e.g. \citealt{Fender2003}, \citealt{Coriat2011}, \citealt{Ecksall2015}). The luminous soft state is radiatively efficient and the accretion can be described by an optically thick, geometrically thin accretion disc (\citealt{Shakura1973}). On the other hand, the hard state is radiatively inefficient and the accretion is due to an advection dominated (ADAF; \citealt{Narayan1995}) or jet dominated acretion flow (JDAF; \citealt{Falke2004}), so that most of the energy output from the accreting BH is in kinetic form. The hypothesis of a state switch connected to the drop in luminosity is consistent also with \cite{Done2007} which show that, as the Eddington ratio decreases, the radiative power of an XRB decreases while the mechanical/kinetic power increases. \cite{Alexander2012} suggested that this is true also for AGN. In this framework, low-luminosity AGN correspond to the hard XRB, while the quasars correspond to the soft state. Different studies showed that the AGN accretion discs can indeed change their state from a classical radiatively-efficient quasar state to a radiatively inefficient accretion flow in a similar way as observed in X-ray binaries. However, since the dynamical and viscous timescales involved in the accretion state changes increase with BH mass, the timescales expected for AGN are significantly longer (e.g. \citealt{Narayan1995}, \citealt{Maccarone2003}, \citealt{McHardy2006}, \citealt{Koerding2006}, \citealt{Schawinski2010b}). As an example, the analogous of a state transition lasting several days in an X-ray binary with a $\sim 10 M_{\rm \odot}$ BH will last $\sim 10^4$ yr in an AGN with a $\sim 10^{7} M_{\rm \odot}$ SMBH. These timescales are consistent with what is seen in IC 2497.

\section{Summary}

We have presented new, deep \chandra X-ray observations of IC 2497. This galaxy is associated with the extended emission line cloud called Hanny's Voorpwerp and hosts a fading AGN. The X-ray data show extended hot gas around the very low-luminosity unobscured AGN ($L_{\rm 2-10 keV}$\,$\sim 10^{40}$\,erg\,s$^{-1}$). Moreover, the data suggest a temperature gradient in the hot gas which may be interpreted as an expanding bubble ($r_{\rm bub} \sim 1$ kpc, $\mbox{{\texttt{E}}$_{\rm bub}$} \sim 10^{54} - 10^{55}$\,erg). One hypothesis is that this bubble is inflated by the AGN that, after changing accretion state and dropping in luminosity, is now in a kinetic mode. As a consequence, the right place to search for and study mechanical AGN feedback could be galaxies hosting fading AGN (i.e. the Voorwerpjes, \citealt{Keel2012}, \citealt{Keel2014}). The idea of an AGN doing mechanical work at the end of its radiatively efficient phase also fits into the general picture of unifying black hole accretion physics from X-ray binaries to AGN. Follow-up deeper \chandra observations will allow us to confirm the presence of the bubble and better constrain its properties and its effect on the host galaxy.

\section*{Acknowledgments}

LFS, KS and MK gratefully acknowledge support from Swiss National Science Foundation Professorship grant PP00P2\_138979/1 and MK support from SNSF Ambizione grant PZ00P2\_154799/1. LS thanks Claudio Ricci and Hans Moritz G\"{u}nther for useful discussions. OIW acknowledges a Super Science Fellowship from the Australian Research Council.

This work is based on observations with the \chandra satellite and has made use of software provided by the \chandra X-ray Center (CXC) in the application packages \ciao and \marxx. The research has made use of NASA's Astrophysics Data System Bibliographic Service.



\appendix

\section{X-ray spectra and {\texttt{Xspec}} fits}\label{app:fit}


\begin{figure*}
\includegraphics[scale=0.5]{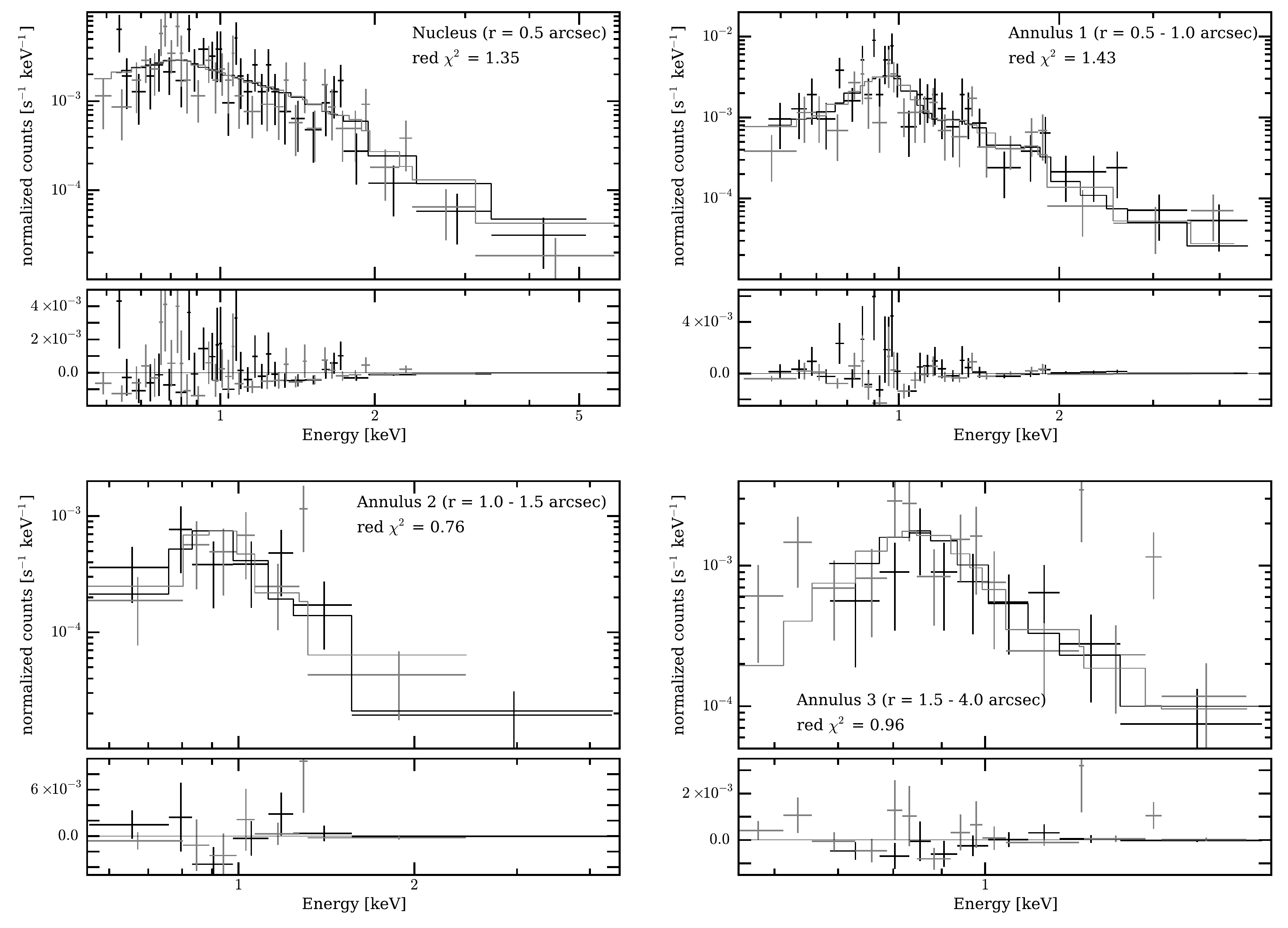}
\caption{\chandra X-ray spectra extracted from the nuclear region and from the three annuli, for ObsID 14381 (black) and ObsID 13966 (grey). The spectra are binned so that each bin contains at least 3 counts. The lines correspond to the best fit model {\texttt{phabs}}$\times$({\texttt{zpow}}+{\texttt{APEC}}), where {\texttt{zpow}} corresponds to the AGN emission and {\texttt{APEC}} to the emission from a hot diffuse gas (see text). The absorption is fixed to the Galactic value. The residuals of the fit are shown in the bottom panel.}
\label{fig:xspec_00_05}
\end{figure*}

\bsp	
\label{lastpage}
\end{document}